# Slow persistent mixing in the abyss

by

**Hans van Haren***

Royal Netherlands Institute for Sea Research (NIOZ) and Utrecht University, P.O. Box 59, 1790  AB Den Burg, the Netherlands.
*e-mail: hans.van.haren@nioz.nl

**Abstract**

Knowledge about deep-ocean turbulent mixing and flow circulation above abyssal hilly plains is important to quantify processes for the modelling of resuspension and dispersal of sediments in areas where turbulence sources are expected to be relatively weak. Turbulence may disperse sediments from artificial deep-sea mining activities over large distances. To quantify turbulent mixing above the deep-ocean floor around 4000 m depth, high-resolution moored temperature sensor observations have been obtained from the near-equatorial southeast Pacific (7°S, 88°W). Models demonstrate low activity of equatorial flow dynamics, internal tides and surface near-inertial motions in the area. The present observations demonstrate a Conservative Temperature difference of about 0.012°C between 7 and 406 meter above the bottom (hereafter, mab, for short), which is a quarter of the adiabatic lapse rate. The very weakly stratified waters with buoyancy periods between about six hours and one day allow for slowly varying mixing. The calculated turbulence dissipation rate values are half to one order of magnitude larger than those from open-ocean turbulent exchange well away from bottom topography and surface boundaries. In the deep, turbulent overturns extend up to 100 m tall, in the ocean-interior, and also reach the lowest sensor. The overturns are governed by internal-wave-shear and -convection. The turbulence inertial subrange is observed to extend into the internal wave frequency band. The associated mixing is not related to bottom friction processes but to internal wave breaking and near-inertial shear. The mixing facilitates long (hours to day) and high (exceeding 100 mab) dispersal of suspended sediments.

**Keywords** Deep-ocean turbulence; abyssal hills plain; high-resolution temperature observations; internal wave mixing; inertial subrange in internal wave band

# 1 Introduction

Motions and concurrent mixing in the ocean play a critical role in the functioning of planet earth. The passage of atmospheric disturbances may cause near-inertial motions in the ocean following geostrophic adjustments in the surface layer (Gill 1982). Near-circular, near-horizontal inertial motions in the ocean have relatively short (ca. 100-m) vertical scales (LeBlond and Mysak 1978). The motions dominate the destabilizing current shear. Across pronounced stabilizing density stratification inertial shear leads to marginal stability and occasional diapycnal exchange that can supply nutrients to the photic zone in shallow seas and upper ocean (van Haren et al. 1999; Alford and Gregg 2001; Lass Prandke and Liljebladh 2003). However, little is known about near-inertial-induced mixing near the bottom in the abyssal ocean, several 1000 m below the surface. Mixing and flow circulation near the seafloor are important for the transfer of benthic derived nutrient fluxes and the resuspension of sediments which may be associated with plumes caused by deep-sea mineral-mining (Miller et al. 2018) and bottom-trawling activities (Puig et al. 2012). Deep-sea mining may be found in relatively flat topographic regions.

In the abyss, away from large steep topography, the water layer above the seafloor is near-homogeneous over a vertical distance that varies from less than 5 m to more than 65 meter above the bottom (hereafter, mab, for short), as shown in relatively rare observational studies (Armi and D'Asaro 1980; Thorpe 1983). Here we define a water layer as near-homogeneous when the vertical, pressure-compensated, density ($\rho$) gradient $d\rho/dz < 10^{-6}$ kg m$^{-4}$, or a buoyancy frequency N $\sim<$ f the inertial frequency (at mid-latitudes). Weakly stratified waters may extend even higher up from the seafloor as observed in shipborne CTD-profile data e.g. from the North-Pacific using more relaxed criteria, e.g. $d\rho/dz < 10^{-5}$ kg m$^{-4}$ (Banyte et al. 2018). The present data overall fall within this weakly stratified range, with a mean of $d\rho/dz \approx$ $8 \times 10^{-6}$ kg m$^{-4}$, over a 400 m depth range.

While the weak stratification is still sufficiently statically stable to support internal gravity waves and suppress, but not block, vertical turbulence exchange, such near-homogeneity over several tens of meters above the seafloor is most likely the result of turbulent mixing. The turbulence is unlikely to be due to friction by current flows over the seafloor, as Ekman dynamics (Ekman 1905) demonstrate this would have a vertical extent of less than 10 mab for eddy viscosities O($10^{-3}$) $m^2 s^{-1}$, which roughly relates with frictional current speeds of about 0.1 $m s^{-1}$. The near-homogeneity is more likely associated with frontal instability, horizontal water mass inhomogeneities (Thorpe 1983), or induced by the breaking of internal, or more precisely inertio-gravity, waves so-called IGW for short. While frontal passages are associated with various mesoscale processes, IGW-breaking over near-flat abyssal plains is less obvious.

The question is why and how IGW become nonlinear and break to potentially drive diapycnal turbulent exchange in the abyss. Can we quantify the variation in abyssal turbulence with time and depth? The question is relevant because the deep-ocean away from large-scale topography is generally considered relatively quiescent in terms of (turbulent) motions (e.g., Polzin et al. 1997). Turbulent motions are considerably weaker in the open-ocean interior than over large-scale topography like, e.g., the Mid-Atlantic Ridge. However, the deep-ocean near-bottom waters over abyssal plains have an enhanced turbulence as determined from temporally limited observations (Thorpe 1983; Polzin et al. 1997). The turbulence may be attributable to the abyssal sea-floor being seldom flat but containing numerous small-scale hills (Armi and D'Asaro 1980). The importance of such topography for IGW-generation has drawn recent scientific interest (e.g., Baines 2007; Morozov 2018). Hitherto, most interest comes from workers undertaking numerical modelling (e.g., Nikurashin et al. 2014; Hibiya Ijichi and Robertson 2017), with few observations.

In this paper high-resolution temperature T-sensor observations are presented from a near-equatorial Pacific abyssal hills site. The observations are used to provide details on IGWs under extraordinary weak stratification conditions. The observations are also used to quantify turbulent mixing and to distinguish shear- and convection-driven processes. As turbulence reaches to within a few meters from the sea-floor its effects on sediment resuspension are hypothesized.

## 2. Data and Methods

Observations were made from the R/V Sonne above the abyssal hills in the Peru Basin DISturbance and reCOLonization (DISCOL) nodule field (https://miningimpact.geomar.de) of the southeast-equatorial Pacific Ocean (Fig. 1). The area around 7°S, 88°W is halfway between the East-Pacific Ridge and the South-American continent, more than 400 km south of the Carnegie Ridge, a potential internal tide source (e.g., de Lavergne et al. 2013). The area is also well away from the equatorial beta plane between latitudes ±2° where energetic trapped waves are found with their specific dynamics (LeBlond and Mysak 1978) including strong near-inertial wave activity near the surface. The observational area is characterized by numerous hills, extending several 100s of meters above the surrounding seafloor at around 4000 m. The average bottom slope is 1.2±0.7°, computed from the lower panel of Fig. 1 using the 1'-resolution version of the Smith and Sandwell (1997) seafloor topography. The slope is similar to within 0.1° of the slope of the Northeast Pacific abyssal plain (van Haren 2018). The slope is about three times larger than that of the Hatteras plain (the area of observations by Armi and D'Asaro 1980) and about three times smaller than that for an area of similar size on the Mid-Atlantic Ridge.

Shipborne SeaBird SBE911plus Conductivity Temperature Depth CTD-profiles were collected once every 5 days within 10 km near 7° 07.2′S, 88° 24.2′ W in a water depth of 4250±20 m, in the period between 13 August and 25 September 2015. The CTD-data were critical as auxiliary information of the temperature-salinity relationship and for post-processing corrections of the moored instrument data. Between 30 July and 28 September Eulerian measurements were made using a taut-wire mooring that was deployed at the above mentioned coordinates.

The lower bound of IGW-frequencies $\sigma$ is determined by the local vertical Coriolis parameter, i.e. the inertial frequency, $f = 2\Omega\sin\varphi$ of the Earth rotational vector $\mathbf{\Omega}$ at latitude $\varphi$. The lower IGW-bound becomes significantly modified to lower sub-inertial frequencies under weak stratification $\propto N^2$, when buoyancy frequency $N < 10f$, approximately. From non-traditional internal wave dynamics equations, minimum and maximum IGW-frequencies are calculated as

$$[\sigma_{min}, \sigma_{max}] = (s -/+ (s^2 - f^2N^2)^{1/2})^{1/2}, \qquad (1)$$

using $2s = N^2 + f^2 + f_h^2\cos^2\gamma$, in which $\gamma$ is the angle to the north ($\gamma = 0$ denoting meridional propagation) and the horizontal component of the Coriolis parameter $f_h = 2\Omega\cos\varphi$ becomes important (e.g., LeBlond and Mysak 1978; Gerkema et al. 2008). From (in)stability analysis of shear working against weak stratification in terms of bulk gradient Richardson number, $N$ is expected to be organized at frequencies $f_h$, $2f_h$, $4f_h$, for nonlinear, linear, nonlinear motions, respectively (van Haren 2008). Locally at the mooring site, $f = 0.181\times10^{-4}$ s$^{-1}$ ($\approx 0.25$ cpd, cycles per day) and $f_h = 1.447\times10^{-4}$ s$^{-1}$ ($\approx 2$ cpd).

The mooring consisted of 2700 N of net top-buoyancy at about 450 m from the seafloor. With maximum current speeds of 0.15 m s$^{-1}$, the buoy did not move more than 0.05 m vertically and 5 m horizontally, as was verified using pressure and tilt sensors attached to single point Nortek AquaDopp acoustic current meters. The mooring held three current meters. These were clamped to the insulated steel mooring cable (0.0063m diameter) at 6, 207 and 408 mab. The upper current meter showed conversion problems and its data are not further considered. A total number of 201 custom-made 'NIOZ4' temperature sensors were taped to the mooring cable at 2.0 m intervals. During the deployment operation, the cable with T-sensors attached was spooled from a custom-made large-diameter drum with separate lanes for sensors and cable (van Haren 2018).

The noise level of the NIOZ4 T-sensor is $< 0.1$ mK, with a precision of $< 0.5$ mK (van Haren et al. 2009). The T-sensors were programmed to sample at a rate of 1 Hz, and synchronized via induction every 4 h, so that every 400-m profile was measured within 0.02 s. Severe constraints were put on the de-spiking, noise levels and drift correction of data because of the small temperature variations observed during the deployment. The data of 33 of the T-sensors (16% of total) was considered deficient due to electronic timing (missed synchronization), battery (power loss), calibration or noise (more than twice the standard noise level) problems and their data was no longer considered and linearly interpolated between data from neighboring functioning T-sensors. The data clean-up biased calculation of turbulence parameters like dissipation rate and diffusivity (see below) to lower values by about 10%.

A strong temperature-density relationship was established (Fig. 2), and after correcting for drift and pressure effects using shipborne CTD-data (Appendix A), the moored T-sensor data allowed us to determine (Thorpe 1977),

$$\text{turbulence dissipation rate } \varepsilon = c_1^2 d^2 N^3, \tag{2}$$

$$\text{vertical eddy diffusivity } K_z = m_1 c_1^2 d^2 N. \tag{3}$$

The variables were calculated following the method of reordering potentially unstable vertical density profiles in statically stable ones, as was originally proposed for shipborne CTD-profile data from a lake by Thorpe (1977). Here, d denotes the displacements between unordered (measured) and reordered profiles and N is computed from the reordered profiles. In (2) and (3), a standard constant value of $c_1 = 0.8$ was used for the Ozmidov/overturn scale factor and $m_1 = 0.2$ for the mixing efficiency, under large Reynolds-number oceanographic conditions (not necessarily laboratory conditions) (e.g., Osborn 1980; Dillon 1982; Oakey 1982; Gregg et al. 2018). These constant values followed as averages over a large number of realizations, with a spread over one order of magnitude. For averaging over all sizes of overturns it is expected that the average includes a mixture of shear-induced Kelvin-Helmholtz instability KHi, and convective columnar turbulence. Both turbulence types occur simultaneously, as columns exhibit secondary shear along the edges and KHi demonstrate convection in their interior core (Li and Li 2004; Matsumoto and Hoshino 2006).

In (3), the validity of $m_1 = 0.2$ is justified after inspection of the temperature-scalar spectral inertial subrange content which is mainly shear-driven (cf. Section 3), and considering the generally long averaging periods over many profiles (>1000). Originally, Thorpe (1977) proposed to consider values of (2) and (3) only for averaging over the size of an entire turbulent overturn in the vertical. The shipborne profiling had a limited temporal resolution to average of the horizontal/time scales of overturns as well. The present moored observations allow for averaging over the largest entire overturns, both in the vertical and in time (over many profiles). For informative purposes graphs will be presented of individual overturning displacements, but quantitative values are only considered after appropriate averaging. In the following, averaging over time is denoted by […], averaging over depth-range by <…>.

Areas of weak and strong turbulence are distinguished using the buoyancy Reynolds number $Re_b = \varepsilon/\nu N^2$ with kinematic viscosity $\nu \approx 1.7 \times 10^{-6}$ m$^2$ s$^{-1}$. Weak turbulence is characterized by $Re_b < 100$ or simply $K_z < 3 \times 10^{-5}$ m$^2$ s$^{-1}$ using (2) and (3).

Because of their sensitivity and frequency the moored T-sensor data are more precise and appropriate than shipborne CTD-data for the application of Thorpe overturning scales for the calculation of turbulence parameters. Hence, most of the concerns raised by e.g. Johnson and Garrett (2004) on Thorpe's method whilst applied to CTD-data are not relevant here. Instead of a single CTD-profile, averaging is performed over $10^3$ to $10^4$ profiles which include at least the buoyancy time scale that exceeds the largest turbulent overturn time-scale, and more commonly the inertial time scale. Moreover, no corrections are needed for 'ship motions' and instrumental-frame flow disturbance, as for CTD-data, because the mooring is not moving more than 0.05 m vertically, on sub-inertial time-scales. A 400 m vertical profile is measured in less than 0.02 s by the moored T-sensors instead of 500 s and obliquely by a shipborne (CTD-)profiler. The noise level of the moored T-sensors is very low, i.e. about one-third of that of the high-precision sensors used in the CTD. Therefore, complex noise reduction as in Piera Roget and Catalan (2002) is not needed for moored T-sensor data.

## 3 Deep-ocean time series observations

### 3.1 Time series and spectral overview

A two-month deployment with the T-sensor array was conducted, but we focus on the last five weeks of the deployment when nearby CTD-observations were available allowing the tuning of the mean background T-profile (Fig. 2). The moored T-observations demonstrate a steady increase in Conservative Temperature (temperature from hereon) with time by about 0.015°C in a month (Fig. 3). The temperature increase is not monotonic, however, and shorter timescale variations are observed. The typical temperature difference between the recordings of two T-sensors placed 378 m apart on the mooring is about 0.01°C, and commensurate the CTD-profiles (Fig. A1). Maximum differences of 0.02°C are recorded around day 247 and minimum values of about 0.003°C around day 262 (Fig. 3). These variations are all smaller than the local adiabatic lapse rate, which is slightly larger than 0.04°C over the 400 m vertical range. Hence, the pressure-effect correction to raw data is required to allow inference of dynamical variations. Within that 400-m range smaller scale steps in temperature stratification are seen in every profile, with typical weakly stratified layer thicknesses of about 50 m. The position above the bottom of such layers varies with time, which is also true near the seafloor, albeit less frequently there.

From the temperature time series data (Figure 3), a persistent periodic signal is not readily discernible, neither on a four-day inertial periodicity, nor on an estimated one-day minimum buoyancy frequency. The lack of persistent periodicity is confirmed by the frequency domain of spectra in Fig. 4. While the kinetic energy of the currents is dominated by a relatively sharp peak at the semidiurnal lunar tidal frequency $M_2$ (Fig. 4a), such a clear periodicity is absent in other parameters. These parameters include the vertically averaged dissipation rate and the large-scale shear, i.e. the vertical (200-m) current difference. The shear shows a broad spectral maximum around f and no significant peaks at other frequencies. As a result, the semidiurnal tidal motions are uniform over at least 200 m and down to 6 m from the bottom, being either barotropic or large-vertical-scale baroclinic. The tidal uniformity in the vertical also implies that Ekman dynamics extend less than 6 mab, because the dominant tidal currents induce the largest bottom-frictional turbulence and an Ekman layer exceeding the height of the lower current meter would result in a tidal current amplitude and phase difference between 6 mab and 200 m higher up. The spectrum of vertically averaged dissipation rate shows a small peak at a frequency just above 2f and near the seafloor only, while being featureless otherwise.

Temperature-spectra are also relatively featureless, with small peaks at diurnal, semidiurnal frequencies and a broad hump towards $N \approx 2f_h$, in the upper range only (Fig. 4b). The temperature-spectra from the lower 100 m of the T-sensor range show a hump around f and 2f. These spectra scale almost uniformly with frequency, largely with $\sigma^{-5/3}$. The slope-value of -5/3 holds for a log-log scale and reflects the passive scalar inertial subrange of dominant shear-induced turbulence (e.g., Tennekes and Lumley 1972). The -5/3-slope is consistent for the range $\sigma \in [f, 80N]$, the latter frequency approximating the spectral roll-off. The spectra from the upper 100-m T-sensor range also follow the -5/3-slope but only for frequencies larger than about twice the maximum buoyancy frequency, $\sigma \in [2N_{max} \approx 8f_h, 80N]$. In the range $[f, 2N_{max}]$ the upper and lower temperature-spectra significantly deviate, the upper one showing a larger energy content with a slope of +2/3 between f and N in the $\sigma^{-5/3}$-scaled log-log plot in Fig. 4b. Hence, the actual spectral slope in that range equals the slope of -1 for open-ocean internal waves (van Haren and Gostiaux 2009). Between N and $N_{max}$ a small peak is noticed at about $3f_h \approx 0.7N_{max}$ which may be related to the displacement of thin layers in the upper range. As the peak lacks the associated π-phase difference, it is not related to a kinematic advection of thin (<2 m, but finite in thickness) interfacial layers passing the sensors (cf. Gostiaux and van Haren 2012), but merely reflects small-scale internal waves peaking at about $0.7N_{max}$, similar to the small peak at about 0.7N.

The variance reduction of lower range temperature with respect to the slope of -5/3 is largest between 0.7 and 1.5 cpd. The 0.7 cpd is just above the frequency of the small peak in lower range dissipation rate (Fig. 4a) and coincides with the reduction in temperature coherence at this point, which is especially noticeable at the larger separation distances >10 m (Fig. 4d). In the upper range well away from the seafloor such a coherence reduction is observed at overall mean N (Fig. 4c). The reduction in the lower range is thus suggested to be associated with the local 'minimum' buoyancy frequency. Although the average estimates of minimum N were twice as high ($N_{min}$ ~1.4 cpd), such averaging reflects the smearing of alternate near-homogeneous and weakly stratified waters. The value of $N_{min,l}$ = 0.7 cpd in the lower range spectra reflects the existence of near-homogeneous waters, as evidenced from time-depth series below. Using this very weak stratification in (1), one finds lower range IGW-bounds of $[\sigma_{min}, \sigma_{max}] = [0.3f, 3N_{min,l}] = [0.08, 2.1]$ cpd (Fig. 4d). The latter frequency coincides with a smaller peak in the coherence. If, as for the upper range, the smaller peak is related to local $0.7N_{max,l}$ this would imply $N_{max,l} \approx 3$ cpd.

The coherence spectra show another unexpected feature, namely a more prolonged higher coherence in the lower 100-m range for the frequency band $\sigma \in$ [N, 80N]. In that frequency band all coherence at all separation scales investigated is higher in the lower range (Fig. 4d), except for the local coherent values at about $0.7N_{max}$ in the upper 100-m range (Fig. 4c). Apparently, the more developed the inertial subrange, the slower coherence is lost with frequency. Or, the more stratified the environment is, the more rapidly coherence is lost outside IGW.

### 3.2 Detailed observations

The observed jump in coherence at local N is suggested here to be related to the mean and minimum N observed in the layers that govern the internal wave motions, as illustrated by an example of relatively weak turbulent motions in Figure 5. For the depicted six-day period, 400-m averaged values are $[<\varepsilon>] = 3\pm2\times10^{-10}$ m$^2$ s$^{-3}$ and $[<K_z>] = 7\pm4\times10^{-4}$ m$^2$ s$^{-1}$, with $[<N>]$ = $2.9\pm0.2\times10^{-4}$ s$^{-1}$. Re$_b$ = 2000. These turbulence values are about an order of magnitude larger than reported for the upper 2000 m of other open-ocean regions (Gregg 1989). With reference to the mean CTD-profile (Fig. A1), temperature and temperature-stratification are seen to vary over a range of time scales, including the inertial scale (the time-scale of variation of near-homogeneous layers near the seafloor and around [4000, 4100] m), approximate diurnal and shorter time scales (mainly in the upper 100 m range). In the plot covering six days the single drift-correction is somewhat problematic because of the relatively long duration. However, the relatively poor correction does not significantly affect the turbulence values for that period. Stratification is found in thin layers of less than 10 m thickness, while largest overturns are found in layers of about 50 to 100 m thickness. Although single overturns are hard to distinguish with smaller overturns occurring in larger ones, apparent single overturns last up to the minimum buoyancy period of about 0.9 days here, e.g. around day 243 at 4000 m, see also the magnification in Figure 6. It is noticeable that the $N_{min}$-overturning is split in two, with single intense overturns lasting about 5 h near the mean buoyancy period. In between, small quasi-convective vertical motions are observed. Both the large-scale overturns before and after the split are genuine overturns, not salinity-compensating intrusions, and they resemble Rankine vortices (Fig. 7). A Rankine vortex has slopes $\frac{1}{2} < z/d < 1$ in the overturn's interior and slopes just exceeding $z/d = 1$ along the sides (van Haren and Gostiaux, 2014). An intrusion model yields interior slopes $z/d > 1$. Although the interior slopes are different for the quasi-convection (Fig. 7b) compared to other larger ones (Fig. 7a,c), they are within the Rankine vortex model range. The occasional erratic appearance in individual profiles, sometimes still visible in the ten-profile means, reflects smaller overturns in larger ones.

The complexity of deep-ocean turbulence is demonstrated by a magnification around the time of maximum overall temperature-difference (Fig. 8). The relatively large temperature-difference is due to isotherms drawing closer together, followed by a rather sudden rise of upper range isotherms and sink of lower range isotherms: quasi mode-2 motion. Shear-driven overturning dominates the upper range which is visible in the roll-up instabilities (Fig. 8a, day 252.4) instead of up- and down-tubes evidencing convection. A near-bottom homogeneous layer is not observed to last long. It is observed to be filled with convective turbulence driven by small-scale internal wave (shear) motions that vary at about 2N (Fig. 8c). Thus, the near-homogeneous layer heights in the lower 100-m range vary considerably, from a few m's to more than 100 m, while the single large interior overturns affect the bottom turbulence as well as shorter period temperature fluctuations, mainly from above. As before, no indications are found of a frictional bottom boundary, which should be predominantly driven by semidiurnal motions, and, if elliptic in form, manifest themselves on a fourth-diurnal (about 6-h) periodicity. In the middle depth-range (Fig. 8b) the isotherm widening through internal wave straining results in a mix of convection and shear-driven overturning.

A 6-h periodicity in near-bottom turbulence is also not observed during a three-day period of relatively frequent and strong turbulent overturning (Fig. 9). For the three-day period, 400-m averaged values are $[<\varepsilon>] = 8\pm5\times10^{-10}$ m$^2$ s$^{-3}$ and $[<K_z>] = 2\pm1\times10^{-3}$ m$^2$ s$^{-1}$, with $[<N>] = 2.9\pm0.2\times10^{-4}$ s$^{-1}$. $Re_b \approx 8000$. These larger than average turbulence values are similar to the ones for the 0.5 day period of Figure 8. The group of large-scale overturns between days 261.5 and 262.5 in Figure 9 is also seen to be driven by a more than 400 m tall straining into an apparent local mode-2 internal wave motion visible as upward displacement of the isotherms in the upper range and downward isotherm displacement in the lower range. The individual overturns again have the duration of about half the mean buoyancy period, while the entire group of overturns lasts about the minimum buoyancy period. Around day 262.5 it seems to be followed by a (series of) frontal motions, trailing highly irregular internal waves representing more likely small-scale overturning reaching throughout the 80-m near-homogeneous layer above the bottom. This shear-driven, quasi-convective turbulence is driven from above and not associated with a frictional process extending upward from the seafloor. As before, no semi- or fourth-diurnal periodicity is observed. The turbulence process from above may be associated with eddy motions that cannot be distinguished from the present observation. However, it also contrasts with frontal bores that whirl up sediment whilst propagating upslope as commonly observed above steep underwater topography. The present fronts are actually seen to have below-average turbulent overturning in their near-bottom interior between days 262.55 and 262.9, z < -4200 m.

Similar larger than average turbulence is occasionally observed in the interior layering (3800-4200 m) (Fig. 10). In the depicted one-day example the stratification is concentrated in a single smaller than 10 m thin layer around 3950 m (cf. also Fig. A1), with nearly 100-m tall overturning alternating on both sides above and below. The observed variation in interior turbulence intensity seems to directly affect internal wave-driven overturning in the lower range which becomes apparent when comparing the correspondence between thin layer isotherm depressions in the upper panels with those of Figure 10d.

## 4 Summary and general discussion

### 4.1 Near-bottom turbulence around equatorial SE-Pacific abyssal hills

In comparison with observations above a hilly abyssal plain in the tropical NE-Pacific (van Haren 2018), the present data from abyssal hills in the SE-Pacific closer to the equator revealed temperature-variations and -stratification that were four times smaller, with variations well below the adiabatic lapse rate. Nevertheless, estimated turbulence values were comparable to within statistical error, being about 20%, larger in the present study. The enhanced turbulence was due to larger sizes of overturns compensating for the weaker stratification at the current study site. While the mean bottom slopes were identical for the two sites, there were nevertheless differences between the observations. The inertial 200-m-shear spread over a much broader frequency band in the present data, while the lower range turbulence peak was of similar width peaking at just above 2f. The latter may reflect the fact that in near-homogeneous layers, near-inertial motions follow paths that are more rectilinear than circular in the horizontal plane because they are increasingly dominated by gyroscopic motions in decreasing stratification (van Haren and Millot 2004). As a result, shear-magnitude, and thus dissipation rate, may manifest itself at 2f, rather than at sub-inertial frequencies as expected for purely circular motions. In both data sets the 200-m-shear removed all semidiurnal periodicity, implying that the tidal current at 6 mab was identical to that higher up, both in amplitude and in phase. As a consequence, the lower current meter (6 mab) was still above the frictional (tidal current) bottom boundary layer as may be inferred from Ekman dynamics for oscillatory flows (Ekman 1905; Maas and van Haren 1987; Simpson et al. 1996). Turbulent mixing in the lower range seem thus predominantly governed by frontal collapse (Armi and D'Asaro 1980; Thorpe 1983) and, in this case possibly more common, by internal wave-induced turbulence from above.

**4.2 Internal wave abyssal mixing**

Occasionally, the internal-wave-induced turbulence appears to be convection dominated, with intrinsic vertical IGW-motions, having speeds O(0.01 m s$^{-1}$) typically observed, probably inducing accelerations exceeding the reduced gravity of the stratification (van Haren 2015). The hypothesized internal-wave convection may be contrasted with the well-known bulk-forcing of acceleration against (or for instabilities with) gravity (e.g., Dalziel 1993). However, on average the turbulence is found to be shear-dominated as inferred from inertial subrange spectral slopes, not only in the more stratified upper range, but especially also in the lower range. In fact, the inertial subrange band is observed to reach into the IGW-band in the lower range. The IGW-band is therefore not exclusively dominated by wave motions. The overlap of the inertial subrange with the IGW-band contrasts with observations from areas where the stratification is larger, e.g. N > 10f, and the buoyancy frequency separates the turbulence- from the IGW-band (e.g., Gregg, 1987). Possibly, the minimum buoyancy frequencies play a role here, being well inside the IGW-band under very weak stratification. The role of $N_{min}$ would match with model observations of stratified grid-turbulence demonstrating a halt of growth after one-quarter buoyancy period (Riley, Metcalf, and Weissman 1981; Itsweire, Helland, and Van Atta 1986), or a frequency of $4N_{min,l} = 3$ cpd here. As a result, the band of moderate coherence may be broadened, spreading over larger scales and more frequencies. The relatively large width of the near-inertial peak reflects a lack of a clear periodic inertial motion source at the site or in time due to the local very weak stratification. While only (meridionally propagating) waves at exactly σ = f can pass any layers of arbitrary stratification between zero and infinity, the observed weak mean N = 0.7 cpd opens an IGW-band of [0.08, 2.1] cpd that includes the above broad near-inertial band also in shear. The wider IGW-band may also be inferred from the broader near-inertial spectral dip in $\varepsilon_l$, which describes the band-spreading of near-inertial source energy into neighbouring bands.

**4.3 The role of rotation on sediment dispersion**

The Earth rotation is thus suggested to play an important role in abyssal turbulent mixing, which is rather slow in its overturns that last quasi-persistently for hours up to a day. While the day-long overturn-persistence is partially due to the proximity of the equator, yielding a four-day inertial period here, it is primarily governed by the weak stratification. However, as the mixing is mainly driven by internal wave motions, re-stratification occurs relatively quickly. As a result, a rather efficient mixing is created as found previously above steeper deep-ocean topography. The observed turbulence dissipation rates are at least tenfold larger than background upper-ocean values (Gregg 1987; Polzin et al. 1997), and about 1000 times the molecular diffusivity. The deep-ocean turbulence will affect bottom resuspension (Hollister and McCave 1984) in an area that may occasionally be calm but is more likely to be stirred a few days later. Artificial reworking as in deep-sea mining will result in long (hours to day) and elevated (larger than 100 mab) resuspension of materials. The study area is not known for extensive geothermal heating in localized thermal vents, which is thus not expected to contribute to the turbulent mixing via convection. However, general geothermal heat fluxes amount approximately 0.1 W m$^{-2}$, depending on the age of the crust (Davies, 2013; Wunsch, 2015). Such heating from below would set-up free convection when stratification is negligible, but it is hardly visible in the present data as it should give: 1. a negative potential temperature gradient which is not visible at all in the mooring and any of the CTD-data (Fig. A1), and 2. A free convection pattern as through DNS-simulations of Rayleigh-Bénard convection in strong rotation providing vertical tubes of convection (e.g., Kunnen et al. 2010) and which are not observed in the present data. This requires further investigations, possibly in combination with high-resolution sediment temperature observations that hitherto hardly demonstrated vertical temperature gradients outside localized vents (Wenzhöfer et al. 2000). Only indirect observational evidence of geothermal heating has been found so far (e.g., Emile-Geay and Madec 2009). In comparison with turbulent mixing above large-scale topography, the present internal wave-induced dissipation rates are 1000 times smaller, but eddy diffusivities are of the same order of magnitude (Polzin et al. 1997).

The spatially one-dimensional observations from this study do not provide information about the source(s) of the internal waves. The inertial motions can be generated by atmospheric disturbances passing over the surface far above, or by local frontal collapse e.g. associated with mesoscale eddies. However, numerical modelling has indicated that the present site shows weak surface input to inertial motions compared to the equatorial region (Rimac et al. 2013). It is expected that not too distant hills may play a role as suggested by numerical modelling efforts, but the present site is not particularly energetic for this source either (Nikurashin et al. 2014; Hibiya Ijichi and Robertson 2017). Similarly, internal tides are particularly weak, e.g. compared to the equatorial Carnegie Ridge region (de Lavergne et al. 2019). Equatorial dynamics including Yanai waves are restricted to ±2° from the equator and do not play a role here (LeBlond and Mysak 1978). Further 3D-observations and modeling may be encouraged to study turbulent overturning details and particular sources in this area. Such local models may then be incorporated in basin-wide near-bottom overturning driven flows (St. Laurent et al. 2001; Ferrari et al. 2016).

**Acknowledgements** I thank M. Laan and L. Gostiaux for their continued discussions plus collaboration in design and construction of NIOZ temperature sensors. I thank the master and crew of the R/V Sonne, J. Greinert and A. Vink for their pleasant contributions to the overboard sea-operations during cruises SO241 and SO242. I sincerely thank my colleagues E. Achterberg, C. Berndt, G. Duineveld, T. Gerkema for critically reading a previous version of the manuscript. Financial support came from the Netherlands Organization for the Advancement of Science (N.W.O.), under grant number ALW-856.14.001 ('Ecological aspects of deep-sea mining' of JPI Oceans project 'MiningImpact').

**Appendix**

In comparison with similar moored abyssal plain NE-Pacific T-observations (van Haren 2018), the present 400-m vertical potential temperature θ difference is only about 0.012°C (Fig. A1) or a factor of four lower and well below the adiabatic lapse rate. Besides a reduction in buoyancy frequency by a factor of two in the present data compared to the NE-Pacific's, the reduction in θ-difference demands a very careful data handling and correction. While the calibration is potentially up to standards with the use of an in-house calibration bath providing <0.1 mK precision, the problem is the correction for sensor bias or drift. Post-processing corrections are imperative for turbulence parameter calculations, especially also the local buoyancy frequency considering (2) and (3). As the instrumentation was used in combination with and after the NE-Pacific deployment, it remained switched on. During the present SE-Pacific deployment the T-sensors were running for at least half a year after programming, i.e. half a year after laboratory calibration.

During post-processing, T-sensor drifts of 1-2 mK/mo are corrected over typical periods of 4 days, the local inertial period, by fitting the average profile to a smooth polynomial profile over the entire vertical range. Such averaging periods need to be at least longer than the buoyancy period to guarantee that the water column is stably stratified by definition (when free convection such as due to geothermal heating is small). In weakly stratified waters as the present observations, the effect of drift is relatively so large that the smooth profile is additionally forced to the smoothed local mean CTD-profile obtained from data collected during the overlapping mooring-period (Fig. A1). The calibrated and drift-corrected T-sensor data are transferred to Conservative (~potential) Temperature (Θ) values (IOC, SCOR and IAPSO 2010), before they are used as a tracer for potential density variations $\delta\sigma_4$, referenced to 4000 dbar, following the constant linear relationship obtained from best-fit data using all nearby CTD-profiles over the mooring period and across the lower 400 m (Fig. 2). As temperature dominates density variations well over the vertical range of moored T-sensors with salinity contributing positively to stability, the relationship's slope is $\alpha = \delta\sigma_4/\delta\Theta = -$

0.261±0.001 kg m$^{-3}$ °C$^{-1}$ (n=5). The resolvable turbulence dissipation rate threshold averaged over a 100-m vertical range is approximately $3\times10^{-12}$ m$^2$ s$^{-3}$.

**Fig. 1**. Bathymetry map of the tropical Southeast Pacific based on the Topo_9.1b -1´ version of satellite altimetry-derived data by Smith and Sandwell (1997). The wagon wheel in the lower panel indicates mooring and CTD positions. Note the different depth ranges between the panels.

**Fig. 2**. Density anomaly referenced to 400 dbar - Conservative Temperature relationship $\delta\sigma_4$ = $\alpha\delta\Theta$ from CTD-data obtained between 3500 and 4200 m near the temperature T-sensor mooring. (a) The data yielding two representative slopes after linear fit are indicated (the mean of five profiles gives the mean relationship-slope $<\alpha>$ = -0.261±0.001 kg m$^{-3}$ °C$^{-1}$). (b) The deviations from the best linear fit and its standard deviation SD.

**Fig. 3**. Time series overview of five weeks of Conservative Temperature from moored T-sensors #5 representing 'upper' (thick graph) and #195 representing 'lower' (thin). The horizontal bars indicate the inertial period (thin line), the approximate 'minimum' buoyancy period (medium) and the mean buoyancy period (thick).

**Fig. 4**. Spectral overview for the period in Fig. 3. (a) Weakly smoothed (10 degrees of freedom, dof) spectra of kinetic energy (middle current meter; thick-dashed graph) and current difference (between lower and middle current meters; thin-dashed). In solid thin and thick the spectra of 150 s sub-sampled time series of 100 m vertically averaged turbulence dissipation rates for lower (7-107 m above the bottom, mab) and upper (307-407 mab) T-sensor data ranges, respectively. The local inertial frequency f, horizontal Coriolis parameter $f_h$ including several higher harmonics, mean buoyancy frequency N $\approx$ 2$f_h$ incl. extent [$f_h$, 2$f_h$], and the semidiurnal lunar tidal frequency $M_2$ are indicated. $N_{max} \approx$ 4$f_h$ indicates the maximum small-scale buoyancy frequency. (b) Strongly smoothed (~1000 dof) spectra of 20 s sub-sampled temperature data representing upper 100 m (thick) and lower 100 m (thin) levels. T-sensor data are low-pass filtered at the roll-off

frequency = 80N, to remove noise. Spectra are scaled with the inertial subrange $\sigma^{-5/3}$, which is a flat horizontal line in the log-log plot as the dashes for reference. (c) Coherence spectra for the upper 100 m of the T-sensor range for all sensor-pairs at the separation distances indicated to the right (larger symbol = thicker line). The 95% significance level is approximately at c = 0.25. (d) As c., but for the lower 100 m of the T-sensor range. The potential minimum buoyancy frequency $N_{min,l}$ for the lower 100 m is inferred from the coherence spectra (see text). Using its value, lower and upper IGW-bounds are computed from (1).

Fig. 5. Six-day time-depth series of moored T-sensor data and turbulence parameter estimates are given for an example of a relatively calm period with some interior turbulent overturning. In all panels the local water depth is at the level of the x-axis and in a.-c. the full 406-m vertical range is shown. (a) Conservative Temperature with 33 out of 201 T-sensor records interpolated because of electronic noise and/or calibration problems. The horizontal bars indicate the inertial period (thin line), the approximate 'minimum' buoyancy period (medium) and the mean buoyancy period (thick). (b) Logarithm of buoyancy frequency from vertically reordered version of data in a. (c) Logarithm of turbulence dissipation rate estimated from data in a. (d) Zoom of lower 100 m of data in a., but with a different temperature range.

Fig. 6. 13-h magnification of a large mid-range overturn from Fig. 5. Note the change to 0.5 mK full-scale of a. The thick horizontal bar in a. indicates the mean buoyancy period of about 6 h. The vertical dashed lines indicate the profiles of Fig. 7.

Fig. 7. Three examples of ten consecutive profiles of displacements (o) and their mean values at given depths (*) for times indicated above and along dashed lines in Fig. 6. The grid consists of two slopes in the d-z plane: ½ (solid lines) and 1 (dashed lines).

**Fig. 8**. Half-a-day example of interior and near-bottom shear- and convection-overturning. The 400-m range is split in three panels (of 155, 155 and 85 m thickness, respectively), each with a different temperature range. The thick horizontal bar in a. indicates the mean buoyancy period.

**Fig. 9**. As Fig. 5, but for a three-day period example of turbulent near-bottom motions. In d. the temperature range is different. The horizontal bars in a. indicate the approximate 'minimum' buoyancy period (medium) and the mean buoyancy period (thick).

**Fig. 10**. As Fig. 9, but for one day period example interior and near-bottom turbulent motions with very thin stratified layers. The thick horizontal bar in a. indicates the mean buoyancy period.

**Fig. A1**. Potential temperature profiles with depth over the lower 400 m above the bottom from CTD-data obtained over a five-week period in August and September 2015 within about 5 km from the moored T-sensors. The mean of the CTD-profiles is given by the circles profile, with its smooth low-pass filtered version in thick solid graph to which the moored data are corrected.

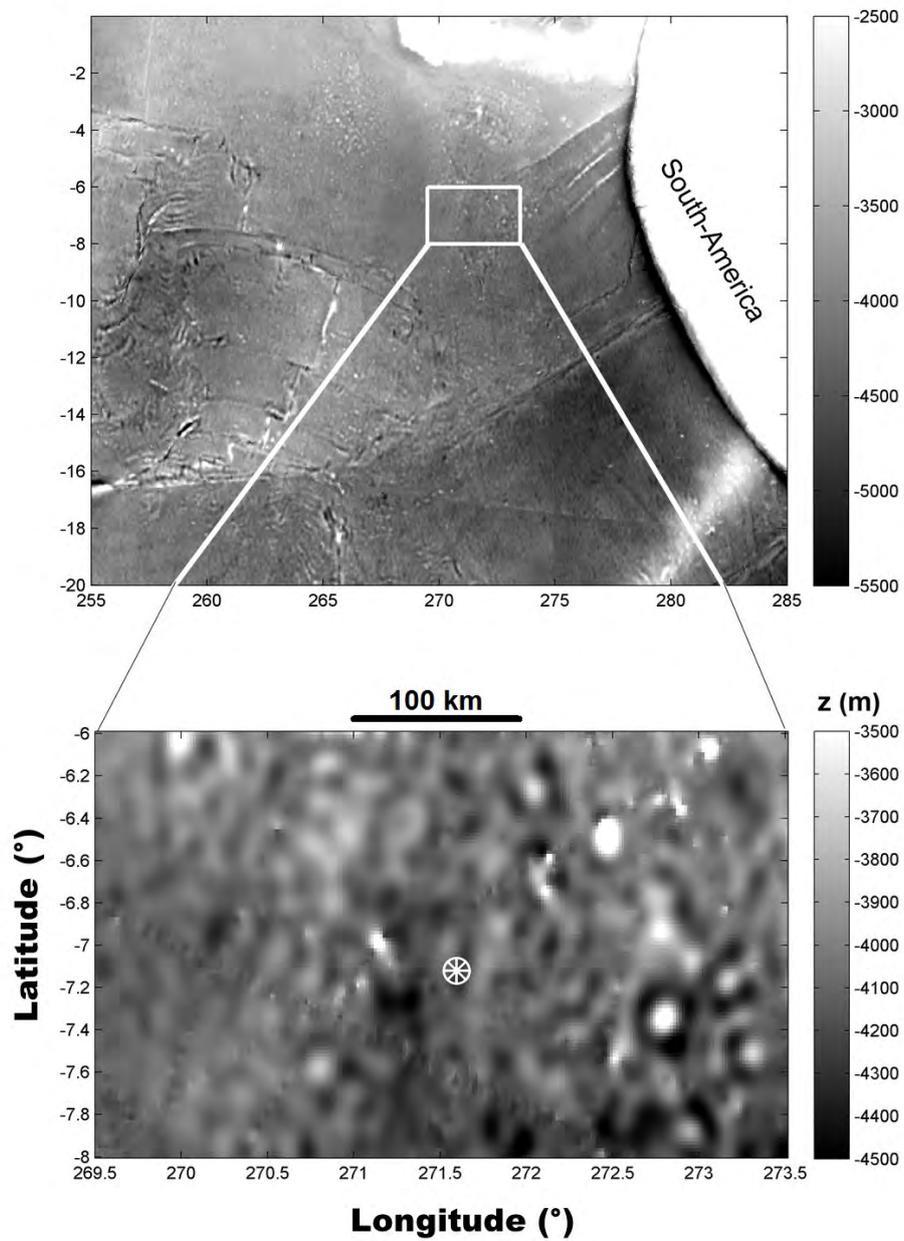

**Fig. 1**. Bathymetry map of the tropical Southeast Pacific based on the Topo_9.1b -1´ version of satellite altimetry-derived data by Smith and Sandwell (1997). The wagon wheel in the lower panel indicates mooring and CTD positions. Note the different depth ranges between the panels.

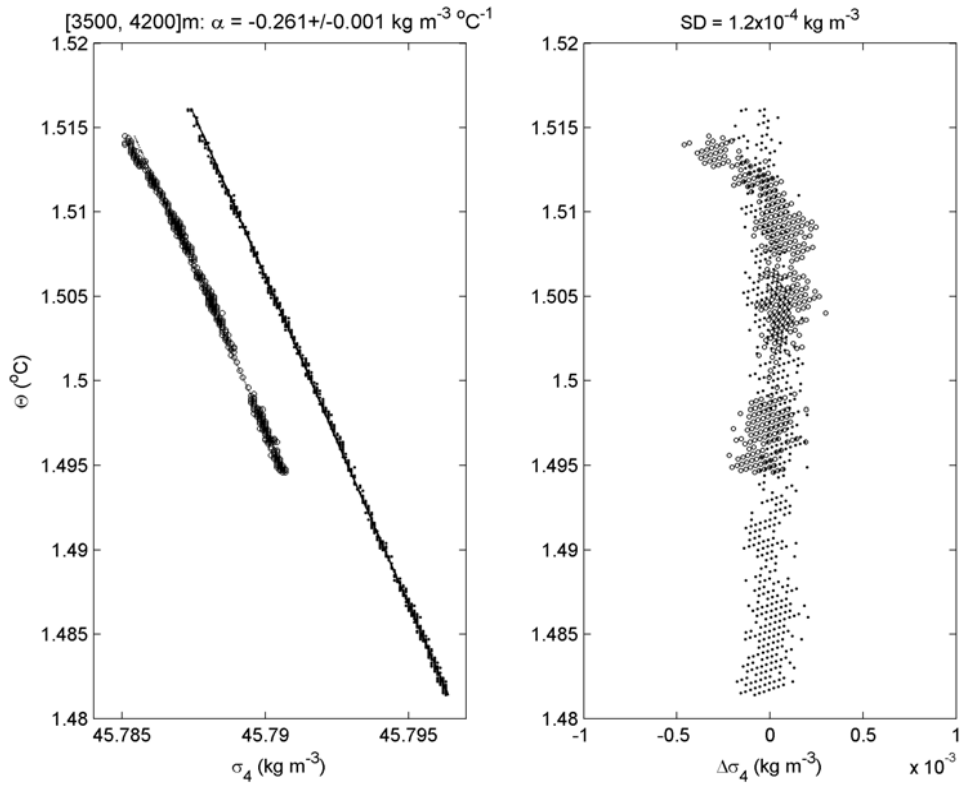

**Fig. 2**. Density anomaly referenced to 400 dbar - Conservative Temperature relationship $\delta\sigma_4 = \alpha\delta\Theta$ from CTD-data obtained between 3500 and 4200 m near the temperature T-sensor mooring. (a) The data yielding two representative slopes after linear fit are indicated (the mean of five profiles gives the mean relationship-slope $<\alpha> = -0.261\pm0.001$ kg m$^{-3}$ °C$^{-1}$). (b) The deviations from the best linear fit and its standard deviation SD.

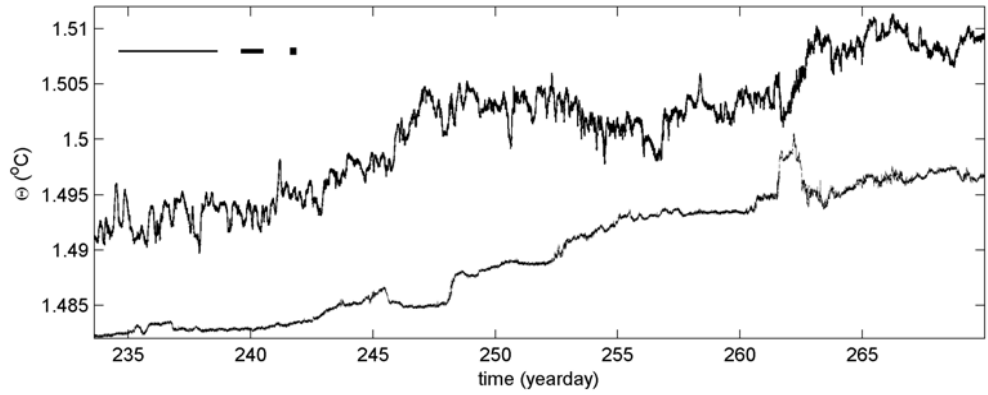

**Fig. 3**. Time series overview of five weeks of Conservative Temperature from moored T-sensors #5 representing 'upper' (thick graph) and #195 representing 'lower' (thin). The horizontal bars indicate the inertial period (thin line), the approximate 'minimum' buoyancy period (medium) and the mean buoyancy period (thick).

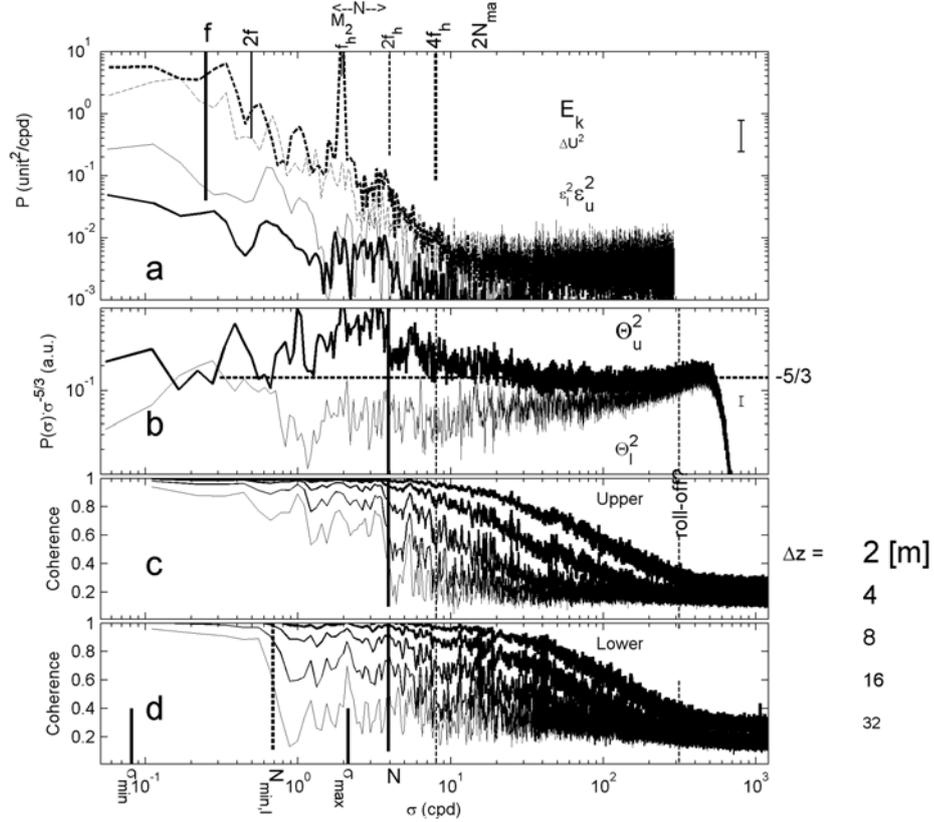

**Fig. 4**. Spectral overview for the period in Fig. 3. (a) Weakly smoothed (10 degrees of freedom, dof) spectra of kinetic energy (middle current meter; thick-dashed graph) and current difference (between lower and middle current meters; thin-dashed). In solid thin and thick the spectra of 150 s sub-sampled time series of 100 m vertically averaged turbulence dissipation rates for lower (7-107 m above the bottom, mab) and upper (307-407 mab) T-sensor data ranges, respectively. The local inertial frequency f, horizontal Coriolis parameter $f_h$ including several higher harmonics, mean buoyancy frequency $N \approx 2f_h$ incl. extend [$f_h$, $2f_h$], and the semidiurnal lunar tidal frequency $M_2$ are indicated. $N_{max} \approx 4f_h$ indicates the maximum small-scale buoyancy frequency. (b) Strongly smoothed (~1000 dof) spectra of 20 s sub-sampled temperature data representing upper 100 m (thick) and lower 100 m (thin) levels. T-sensor data are low-pass filtered at the roll-off frequency = 80N, to remove noise. Spectra are scaled with the inertial subrange $\sigma^{-5/3}$, which is a flat horizontal line in the log-log plot as the dashes for reference. (c) Coherence spectra for the upper 100 m of the T-sensor range for all sensor-pairs at the separation distances indicated to the right (larger symbol = thicker line). The 95% significance level is approximately at c = 0.25. (d) As c., but for the lower 100 m of the T-sensor range. The potential minimum buoyancy frequency $N_{min,l}$ for the lower 100 m is inferred from the coherence spectra (see text). Using its value, lower and upper IGW-bounds are computed from (1).

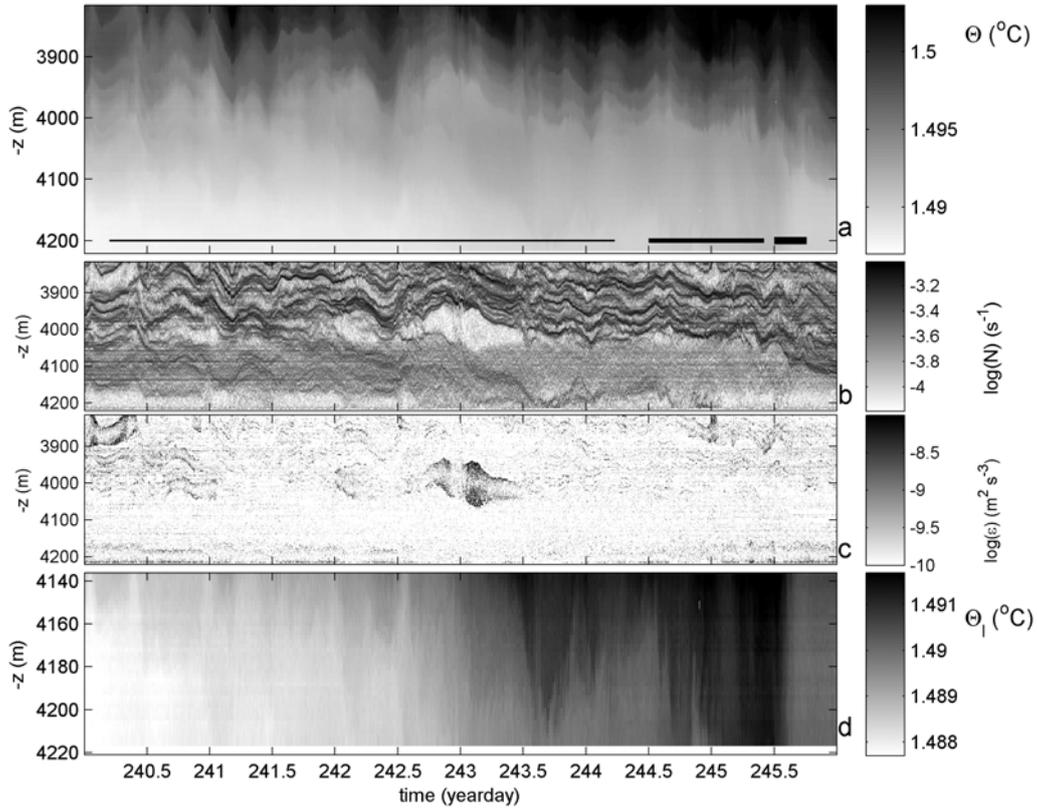

**Fig. 5**. Six-day time-depth series of moored T-sensor data and turbulence parameter estimates are given for an example of a relatively calm period with some interior turbulent overturning. In all panels the local water depth is at the level of the x-axis and in a.-c. the full 406-m vertical range is shown. (a) Conservative Temperature with 33 out of 201 T-sensor records interpolated because of electronic noise and/or calibration problems. The horizontal bars indicate the inertial period (thin line), the approximate 'minimum' buoyancy period (medium) and the mean buoyancy period (thick). (b) Logarithm of buoyancy frequency from vertically reordered version of data in a. (c) Logarithm of turbulence dissipation rate estimated from data in a. (d) Zoom of lower 100 m of data in a., but with a different temperature range.

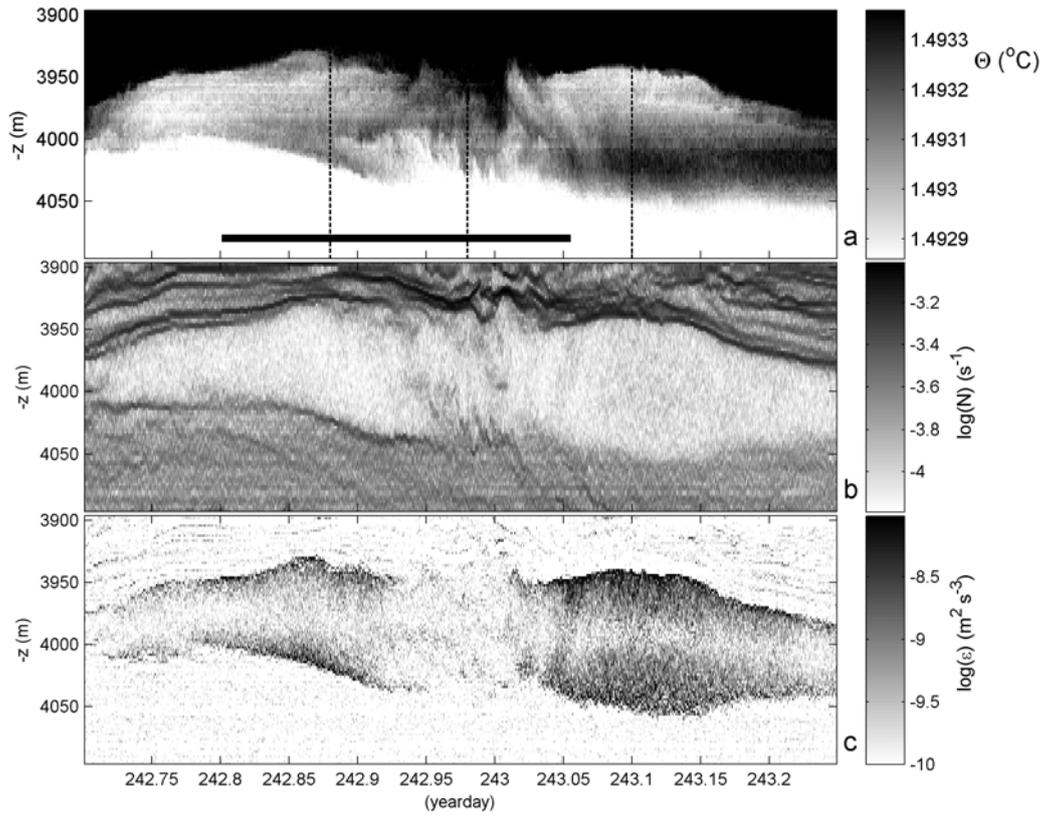

**Fig. 6**. 13-h magnification of a large mid-range overturn from Fig. 5. Note the change to 0.5 mK full-scale of a. The thick horizontal bar in a. indicates the mean buoyancy period of about 6 h. The vertical dashed lines indicate the profiles of Fig. 7.

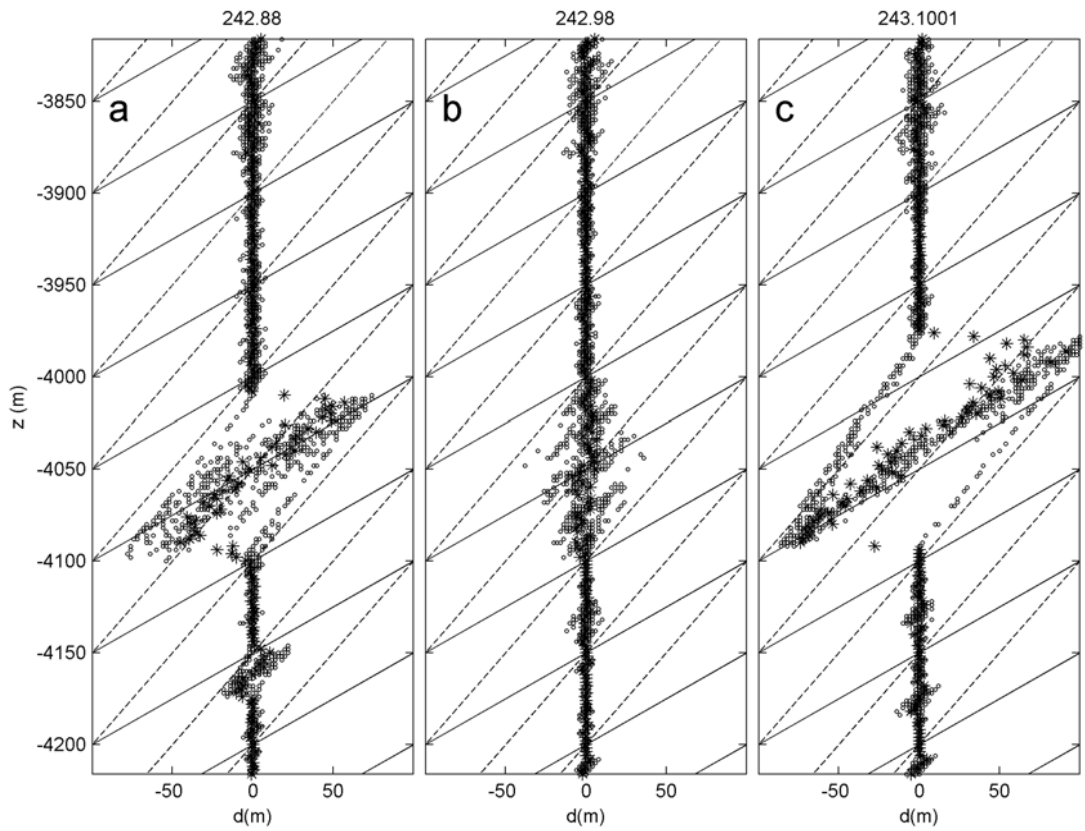

**Fig. 7**. Three examples of ten consecutive profiles of displacements (o) and their mean values at given depths (*) for times indicated above and along dashed lines in Fig. 6. The grid consists of two slopes in the d-z plane: ½ (solid lines) and 1 (dashed lines).

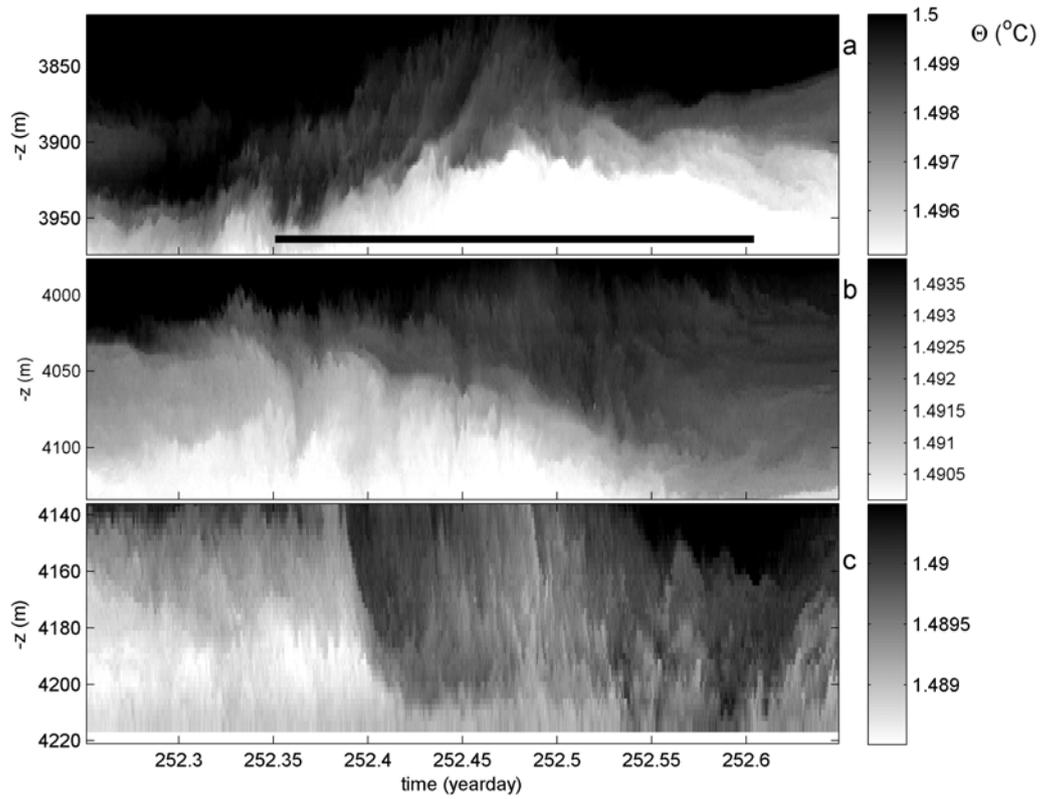

**Fig. 8**. Half-a-day example of interior and near-bottom shear- and convection-overturning. The 400-m range is split in three panels (of 155, 155 and 85 m thickness, respectively), each with a different temperature range. The thick horizontal bar in a. indicates the mean buoyancy period.

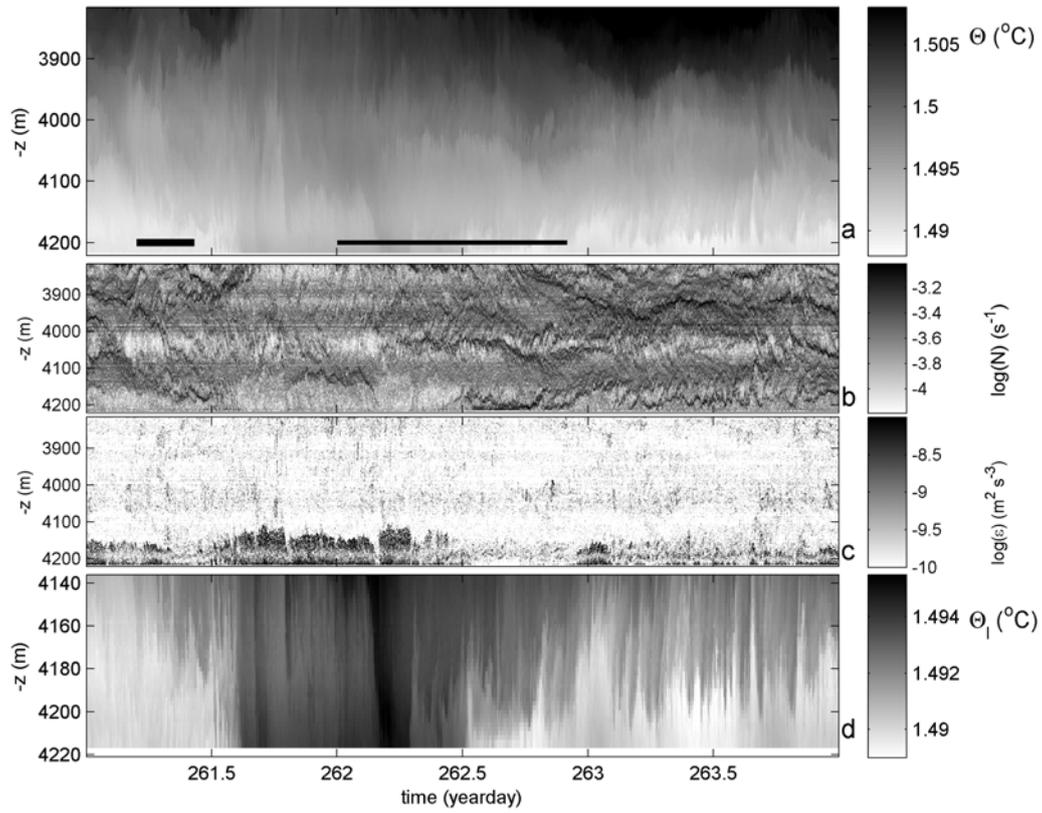

**Fig. 9**. As Fig. 5, but for a three-day period example of turbulent near-bottom motions. In d. the temperature range is different. The horizontal bars in a. indicate the approximate 'minimum' buoyancy period (medium) and the mean buoyancy period (thick).

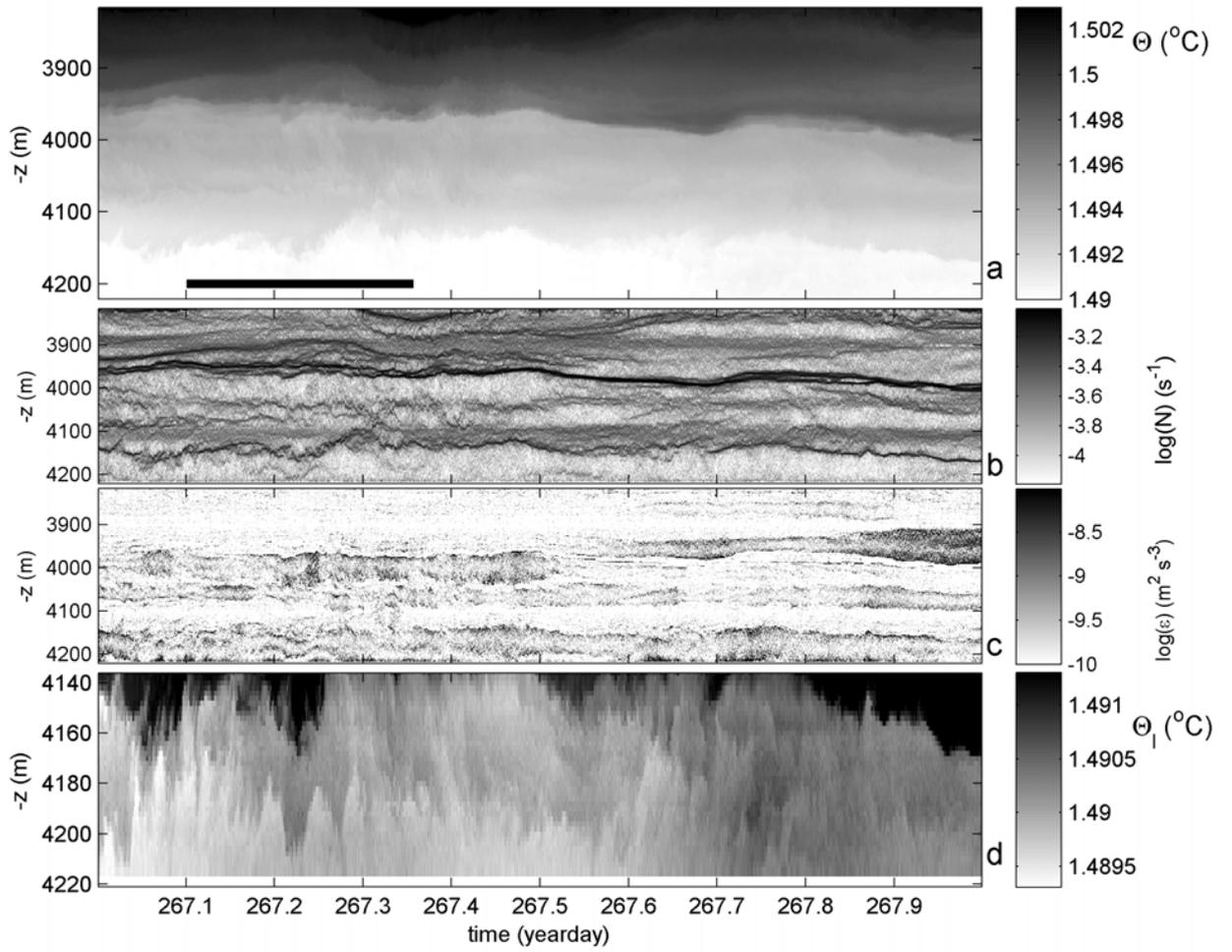

**Fig. 10**. As Fig. 9, but for one day period example interior and near-bottom turbulent motions with very thin stratified layers. The thick horizontal bar in a. indicates the mean buoyancy period.

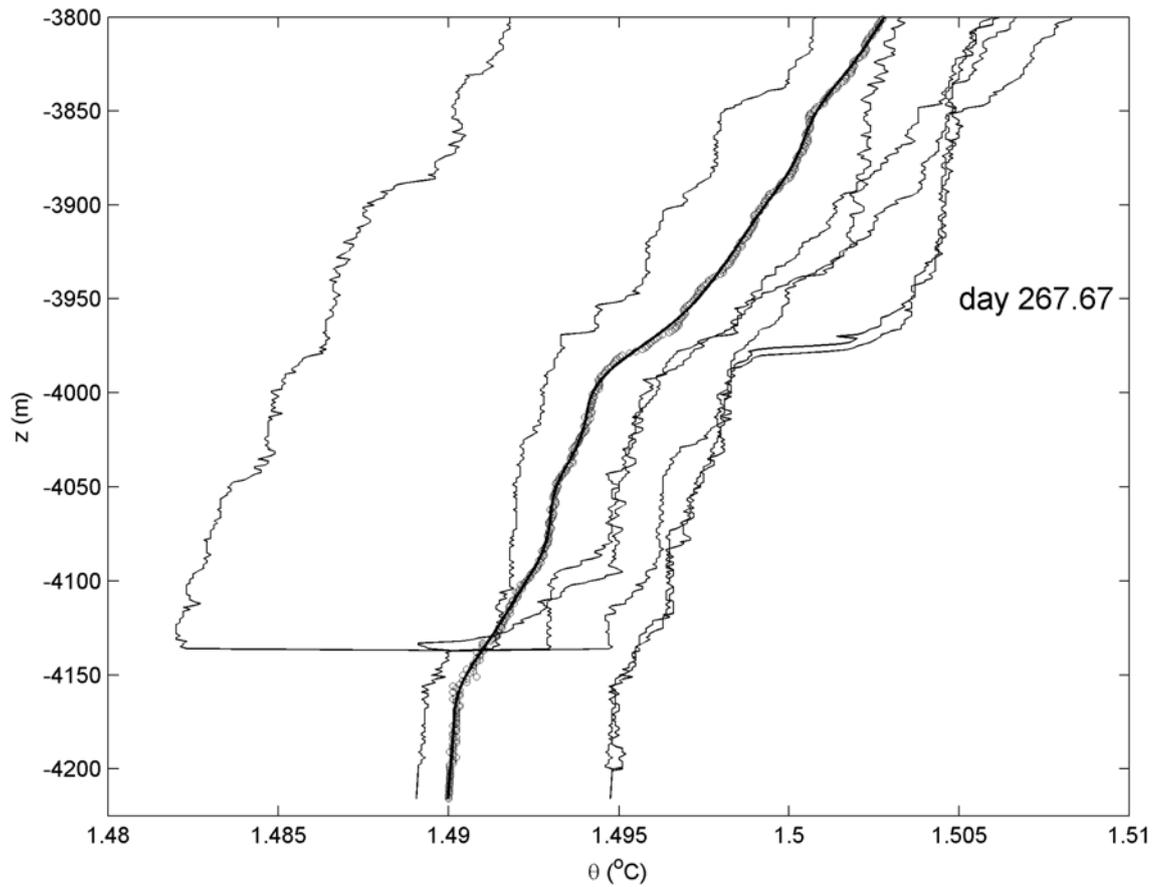

**Fig. A1**. Potential temperature profiles with depth over the lower 400 m above the bottom from CTD-data obtained over a five-week period in August and September 2015 within about 5 km from the moored T-sensors. The mean of the CTD-profiles is given by the circles profile, with its smooth low-pass filtered version in thick solid graph to which the moored data are corrected.